\documentclass{emulateapj}
\usepackage{times, graphicx}

\newcommand{\cjaa}{ChJAS}

\slugcomment{}

\shorttitle{Chromospheric Nanoflares}
\shortauthors{D.B. Jess et al.}

\begin{document}

\title{Nanoflare Activity in the Solar Chromosphere}

\author{D. B. Jess$^{1}$, M. Mathioudakis$^{1}$, P. H. Keys$^{1,2}$}
\affil{$^{1}$Astrophysics Research Centre, School of Mathematics and Physics, 
Queen's University Belfast, Belfast BT7 1NN, UK}
\affil{$^{2}$Solar Physics and Space Plasma Research Centre (SP$^{2}$RC), 
University of Sheffield, Hicks Building, Hounsfield Road, Sheffield S3 7RH, UK}
\email{d.jess@qub.ac.uk}

\begin{abstract}
We use ground-based images of high spatial and temporal resolution to 
search for evidence of nanoflare activity in the solar chromosphere. 
Through close examination of more than $1 \times 10^{9}$~pixels in the 
immediate vicinity of an active region, we show that the distributions 
of observed intensity fluctuations have subtle asymmetries. 
A negative excess in the intensity fluctuations indicates that more 
pixels have fainter-than-average intensities compared with those 
that appear brighter than average. 
By employing Monte Carlo simulations, we reveal how the negative excess 
can be explained by a series of impulsive events, coupled with exponential 
decays, that are fractionally below the current resolving limits of 
low-noise equipment on high-resolution ground-based observatories. 
Importantly, our Monte Carlo 
simulations provide clear evidence that the intensity asymmetries 
cannot be explained by photon-counting statistics alone. A comparison 
to the coronal work of \citet{Ter11} suggests that nanoflare activity in the 
chromosphere is more readily occurring, with an impulsive event occurring 
every $\sim$$360$~s in a $10{\,}000$~km$^{2}$ area of the chromosphere, 
some $50$ times more events than a comparably sized region of the corona. 
As a result, nanoflare activity in the chromosphere is likely to play an important 
role in providing heat energy to this layer of the solar atmosphere.
\end{abstract}

\keywords{methods: numerical --- Sun: activity --- 
Sun: chromosphere --- Sun: flares}

\section{Introduction}
\label{intro}
Magnetic reconnection is a common phenomenon within the solar 
atmosphere. Its presence is often observed through explosive 
events such as solar flares, where extreme localized heating is generated 
through the conversion of magnetic energy \citep{Pri86, Pri99}. Large-scale 
flare events can be dramatic, often releasing in excess of $10^{31}$~ergs 
of energy during a single event. However, the relative rarity of these 
phenomena means that they cannot provide the necessary sustained heating 
to maintain the multi-million degree temperatures observed in the outer solar 
atmosphere. 
Instead, it has been suggested that nanoflares, 
with an energy of approximately $10^{24}$~ergs, 
may occur with such regularity in the vicinity of active 
regions that they can provide a basal background heating \citep{Parker88}. 
Previous work on nanoflare heating has focused on coronal 
observations and modelling, with spectroscopic techniques used to 
investigate the scaling between the emission measure and the 
temperature of coronal plasma 
\citep[e.g.,][]{Kli01, Bra12}. 
These results tentatively suggest that nanoflare heating may be 
responsible for a significant fraction of the energy deposited in the 
outer solar atmosphere. However, the reliability of these techniques 
hinge on the accuracy of the emission measure diagnostics as well 
as the number of optically-thin magnetic strands superimposed along an 
observational line-of-sight. Indeed, recent work by 
\citet{Cir13}, who employed the high-resolution 
sounding-rocket imager Hi--C, found a wealth of fine-scale coronal 
structuring that is below the diffraction limit of current space-based 
coronal observatories during the instrument's 5-minute flight. 
To avoid the emission measure sensitivities to local plasma temperatures, 
\citet{Ter11} employed direct imaging techniques and undertook a 
statistical study utilising X-ray data collected by the 
X-Ray Telescope \citep[XRT;][]{Gol07} onboard Hinode, to 
investigate whether the analysis of millions of pixels as a single collective 
could refute or verify the presence of nanoflares in the Sun's corona. 
The authors detected a small asymmetry in the measured intensity 
fluctuations, which they interpreted as the signature of cooling plasma 
induced by a sequence of impulsive reconnection events. 
Consequently, \citet{Ter11} suggested that 
nanoflares are a universal heating process within solar active regions. 
Unfortunately the signal-to-noise, frame rate, and spatial resolution of 
the observations were not sufficient to unequivocally determine the 
presence of nanoflare activity and evaluate the specific role they play 
in the heating of the Sun's outermost atmosphere.

\begin{figure*}
\epsscale{1.0}
\plotone{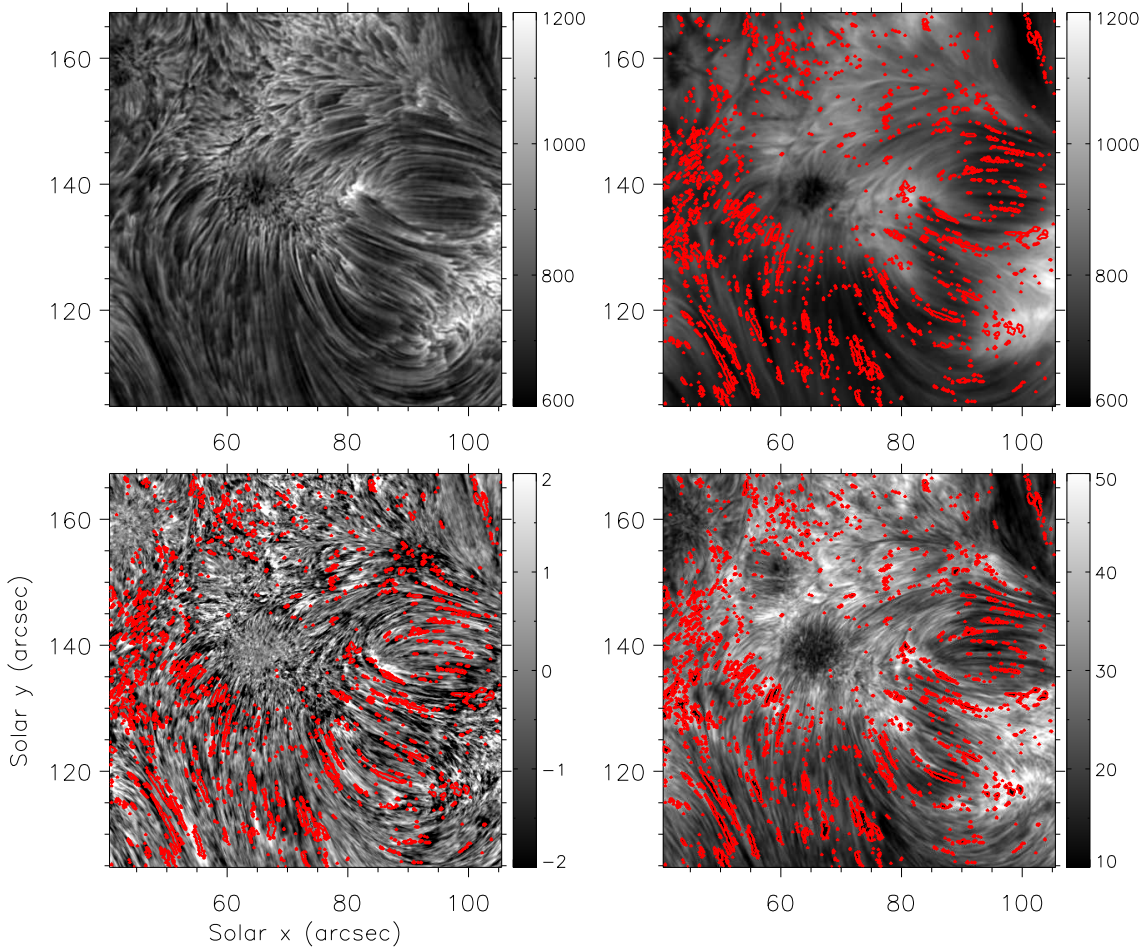}
\caption{An H$\alpha$ core (upper-left) snapshot, acquired at 
$17$:$52$~UT on 2011 December 10. A time-averaged H$\alpha$ 
core image (upper-right) is generated by averaging all $4040$~individual 
images acquired during the 2~hour duration of the data set. Colorbars beside 
the two upper panels denote the image intensities in DN{\,}s$^{-1}$. 
The lower-left panel displays the time-averaged pixel medians (normalised to their 
standard deviation, $\sigma$, and artificially saturated to assist the clarity of 
small-scale features), while the lower-right panel displays the standard 
deviations (in DN{\,}s$^{-1}$) for the entire field-of-view. 
Red contours outline regions excluded 
from analysis, and the axis scales are in heliocentric coordinates, where 
$1{\arcsec}\approx725$~km. An animation of this figure is 
available in the online journal.
\label{images}}
\end{figure*}

While the majority of recent nanoflare studies have been dedicated 
to coronal emission, it is the solar chromosphere that provides more 
tantalising prospects for rapid advancements in solar physics. 
In the current era we have numerous observational 
facilities at our disposal that provide a wealth of 
high spatial, spectral and temporal resolution chromospheric 
observations. Such observatories include the ground-based 
Dunn and Swedish Solar Telescopes equipped with the 
ROSA \citep{Jes10c} and CRISP \citep{Sch08} 
instruments, respectively, alongside the space-based 
Hinode satellite. Of particular note is the recently launched 
Interface Region Imaging Spectrograph \citep{DeP14}, 
which aims to bridge the gap between traditional optical 
observations of the chromosphere and their 
corresponding coronal EUV counterparts.
Even though the chromosphere is only heated to a few thousand 
degrees above the corresponding photospheric layer, the high 
densities found within the chromosphere require $2$--$3$ times 
more energy input to maintain its temperature when compared to 
the multi-million degree coronal plasma \citep{Wit77, And89}.
Recent work 
has revealed that flare signatures can be contained within the 
chromospheric layer, supporting a wealth of low-lying impulsive 
events including Ellerman bombs and 
H$\alpha$ microflares \citep{Din99, Che01, Jes10a, Nel13}. 
Furthermore, flares emit most of their radiative signatures in 
the optical and UV portion of the electromagnetic spectrum 
\citep{Nei89, Woo06}, and as a result, 
the chromosphere is also the primary energy loss region 
associated with such impulsive events \citep{Fle11}. 
Thus, the chromosphere 
presents an ideal, and previously unexplored observational 
platform to investigate the role nanoflare activity plays in the 
heating of the Sun's dynamic atmosphere.

In this paper, we utilise high spatial and temporal resolution 
observations of the solar chromosphere to investigate 
whether nanoflare activity can be detected in a relatively 
quiet active region, devoid of any large scale magnetic activity. 
We employ a collection of techniques previously used by 
\citet{Ter11}, to study the statistics of chromospheric intensity 
fluctuations, and ultimately relate the analysis of millions of 
individual pixels to the detection of nanoflare events.

\section{Observations}
\label{obs}
The observational data presented here are part of a sequence obtained during 
$17$:$51$ -- $19$:$51$~UT on 2011 December 10, with the Dunn Solar 
Telescope (DST) at Sacramento 
Peak, New Mexico. The newly-commissioned 
Hydrogen-Alpha Rapid Dynamics camera \citep[HARDcam;][]{Jess12} 
imaging system was employed to image a 
location surrounding active region NOAA 11372, positioned at heliocentric co-ordinates 
($71${\arcsec}, $134${\arcsec}), or N$07.6$W$04.2$ in the conventional 
heliographic co-ordinate system. 
HARDcam observations employed a $0.25${\,}{\AA} filter centered on the 
H$\alpha$ line core (6562.8{\,}{\AA}), and utilized a spatial 
sampling of $0{\,}.{\!\!}{\arcsec}138$ per pixel, providing a 
field-of-view size of $71{\arcsec}\times71{\arcsec}$. 
During the observations, high-order adaptive optics \citep{Rim04} 
were used to correct wavefront deformations in real-time. The acquired images were 
further improved through speckle reconstruction algorithms \citep{Wog08}, 
utilizing $35 \rightarrow 1$ restorations, 
resulting in a reconstructed cadence of $1.78$~s. 
Atmospheric seeing conditions remained excellent 
throughout the time series. However, to ensure accurate co-alignment, 
narrowband HARDcam images were Fourier co-registered and 
corrected for atmospheric warping through the application of destretching 
vectors established from simultaneous broadband reference images 
\citep{Jes07}. Sample images, incorporating all 
image processing steps and including a time-averaged reference image, 
can be viewed in Figure~\ref{images}.

\section{Data Analysis \& Interpretation}
\label{analy}
\subsection{Observational Time Series}
During the two hour duration of the observing sequence, no large scale eruptive 
phenomena (GOES A-class or above) were observed from the active region 
under investigation. Examination of a time-lapse movie of HARDcam H$\alpha$ 
images revealed no large-scale structural re-configurations or periodic 
motions associated with spicules \citep{Jes12b}, fibrils \citep{Mor11, Mor12}, or 
mottles \citep{Kur12}. This magnetically ``locked'' configuration is verified 
through examination of the time-averaged 
H$\alpha$ image displayed in the upper-right panel of Figure~{\ref{images}}. 
Fine-scale structuring can still 
readily be observed, even after the images have been averaged over 
the entire $2$~hour ($4040$~frames) duration of the dataset, indicating 
a rigid chromospheric canopy with little-to-no periodic motions which would 
have caused intrinsic blurring in the time-averaged image.  

Following 
the methodology of \citet{Ter11}, our time series was subjected to data cleaning 
procedures, including the removal of pixels with excessively low count rates, 
those affected by macroscopic 
\citep[i.e., H$\alpha$ microflare;][]{Jes10a}
brightenings, and those 
demonstrating slow intensity variations due to the displacement or drift of 
structures within the field-of-view. 
The signal-to-noise of the time series was high, mostly attributed to the low 
dark current provided by the Peltier-cooled back-illuminated 
CCD \citep{Jes10c}. An average count rate of 
$785$~DN{\,}s$^{-1}$ was present, with the darkest parts of the 
sunspot umbra and the brightest regions of the chromospheric canopy 
remaining above $550$~DN{\,}s$^{-1}$ and below $1550$~DN{\,}s$^{-1}$, 
respectively. The time series signal-to-noise ratio, $S/N$, can be calculated as 
$S/N = \mu / \sigma_{N}$, where $\mu$ is the signal mean and $\sigma_{N}$ is 
the standard deviation of the noise under normal observing conditions 
\citep{Sch00, Jes12c}. The image noise will contain contributions from both 
the detector readout and pixelised photon statistics, with the standard 
deviation of the former $\sigma_{d}\approx4.3$~DN{\,}s$^{-1}$ (derived from 
$4000$ consecutive dark frames), while the standard deviation of the 
latter equated as $\sigma_{p}=\sqrt{n-\sigma_{d}^{2}}$, where $n$ is the individual 
pixel counts in DN{\,}s$^{-1}$. The total noise contribution is found from 
$\sigma_{N}=\sqrt{\sigma_{d}^{2}+\sigma_{p}^{2}}=\sqrt{n}$, providing an average 
observational standard deviation 
$\approx$$28$~DN{\,}s$^{-1}$
(lower-right panel of Figure~{\ref{images}}). Thus, 
to estimate the range of $S/N$ values found in the observations, the extreme 
detector counts of $550$~DN{\,}s$^{-1}$ and $1550$~DN{\,}s$^{-1}$ 
provide a signal-to-noise range of 
$S/N \approx 23-39$.
As a result of the high signal-to-noise values, no 
pixels within the field-of-view were discarded 
on the basis of poor count statistics. To remove macroscopic variations, a 
linear fit was performed on individual pixel lightcurves, with pixels removed 
which had intensities reaching or exceeding $150$\% of the best-fit line at 
any time. These accounted for $\approx$$0.34$\% ($886$~pixels) of the total. 

Finally, pixels which displayed long-term intensity variations, caused by either 
the displacement or drift of the structures present, were removed from the 
field-of-view. To do this, we assumed that if the fluctuations around the 
linear fit were completely random, and followed a binomial distribution with 
$0.5$ probability of crossing the line-of-best-fit at any time, then the number of crossings 
due to structural displacements and/or drifts should be smaller than the standard 
deviation of the binomial distribution \citep{Ter11}. As a result, 
all pixels were removed which 
had intensities that crossed the line-of-best-fit less than $\sqrt{(m-1)}/2$ 
times, where $m$ is the number of data points. Thus, for our $2$~hour 
time series incorporating $4040$ individual time stamps, all pixels were 
discarded where their respective lightcurves crossed the best-fit line 
less than $32$~times. These pixels cover $\approx$$0.34$\% ($891$~pixels) 
of the total field-of-view. However, observations acquired in H$\alpha$ may also 
capture dynamic periodic phenomena, such as spicules, mottles, and 
fibrils. As mentioned above, a time-lapse movie of the observations 
revealed no periodic transverse motions, implying a rigid magnetic 
configuration is present. However, to remove pixels that contain even the most 
subtle oscillating structures, we calculated the 
upper crossing threshold which would arise as a result of features 
oscillating with the lowest transverse periodicities measured in previous 
H$\alpha$ studies. A structure oscillating with a transverse periodicity of 
$\approx$$70$~s \citep{Kur12} would cross the best-fit line twice during a complete 
oscillation cycle. Thus, over the $2$~hour ($\approx$$7200$~s) 
duration of the dataset, one would expect $\approx$$205$ crossings 
of the best-fit line. This is a gross overestimate, as it assumes that the 
lifetime of an oscillating H$\alpha$ feature is longer than the $2$~hour 
duration of the dataset. Nevertheless, by neglecting pixels within our 
field-of-view which cross the best-fit line less than $205$ times, we only 
discard $\approx$$2.98$\% ($5477$~pixels). 

\begin{figure*}
\epsscale{0.95}
\plotone{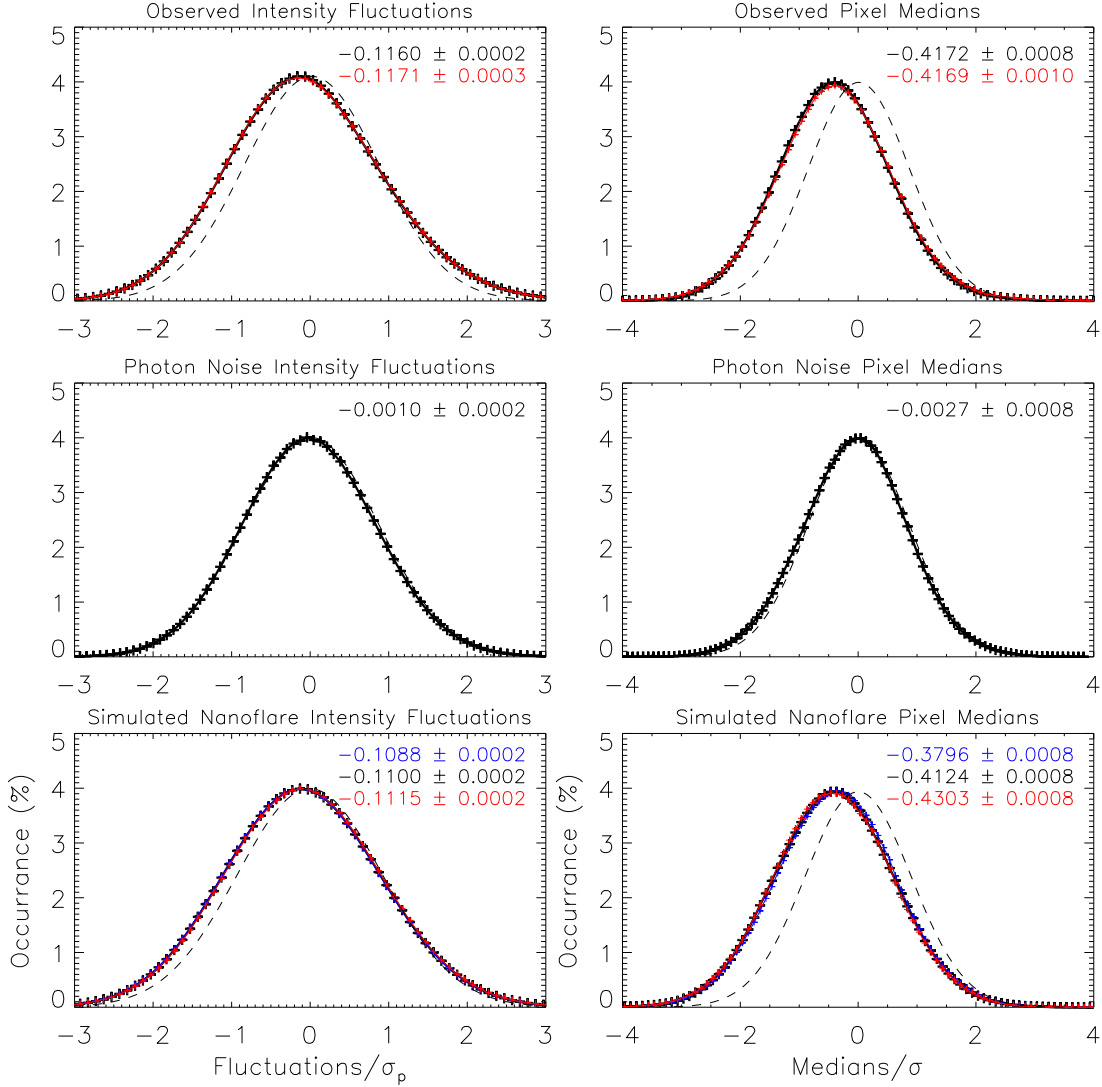}
\caption{The upper-left panel displays the distribution of observed 
H$\alpha$ pixel intensity fluctuations for the entire field-of-view 
(red line) and for those which pass the threshold 
criteria outlined in \S~\ref{analy} (black line), normalized to the 
photon noise, $\sigma_{P}$. 
The upper-right panel shows the distribution of median fluctuations, 
computed individually for each pixel within the entire field-of-view 
(red line) and those which pass the threshold criteria (black line), 
both of which are normalized to their standard deviation, $\sigma$. The remaining panels are 
identical to the upper distributions, but created for synthetic data sets 
which display fluctuations based entirely upon Poisson noise statistics (middle) 
and Monte Carlo simulations of nanoflare activity 
($A = 20$~DN{\,}s$^{-1}$, $dt = 360$~s; lower). The blue, black and red 
lines in the lower panels represent $e$-folding times of 
$\tau = 37${\,}s, $51${\,}s and $65${\,}s, respectively. For each distribution, 
a Gaussian centered on zero with unit width is displayed as a dashed 
line for reference, while the centroid offsets for individual distributions are displayed 
in the upper-right corner of each panel. 
\label{distributions}}
\end{figure*}

Following the rigorous data cleaning, more than $96.6$\% 
of the total number of pixels remained, providing in excess of 
$1.02 \times 10^{9}$ individual pixels. Regions removed from subsequent 
analysis are contoured in red in Figure~{\ref{images}}. Intensity 
fluctuations, $dI$, of the remaining pixels were computed similarly to 
\citet{Ter11},
\begin{equation}
dI(x,y,t) = \frac{I(x,y,t) - I_{0}(x,y,t)}{\sigma_{P}(x,y,t)} \ ,
\end{equation}
where $I(x,y,t)$ is the count rate (DN{\,}s$^{-1}$), 
$I_{0}(x,y,t)$ is the value of the linear fit, and 
$\sigma_{P}(x,y,t)$ is the photon noise estimated as the standard deviation 
of the pixel lightcurve with respect to the linear fit, acquired at the 
spatial position [$x$, $y$] and time $t$. The slopes of best-fit lines 
for each pixel are very small ($0 \pm 0.12$), and as noted by 
\citet{Ter11}, show no preference for increasing or decreasing intensities. 
Due to each lightcurve being 
normalised to its own respective best-fit line, a more statistically significant 
distribution can be obtained by including fluctuations over the entire 
field-of-view which are not removed by the process of data cleaning. 
By definition, the mean fluctuation for each pixel is $0$. 
However, the upper-left panel of Figure~{\ref{distributions}} clearly 
displays a negative excess in the intensity fluctuations 
(normalised to $\sigma_{P}$), indicating more 
pixels have fainter-than-average 
intensities compared with those that appear brighter than average. 
Averaged over all pixels that passed the data cleaning criteria, the 
measured median fluctuation is $-0.1160 \pm 0.0002$. 
The upper-right panel of Figure~{\ref{distributions}} displays the 
distributions of the median values themselves (normalised to their 
standard deviation), computed individually 
at each pixel. Again, there is a preference for the median value 
to be negative with respect to the mean, indicating the presence 
of a widespread and real statistical phenomenon. Here, the measured 
median average for all pixels that passed the data cleaning criteria 
is $-0.4172 \pm 0.0008$. These effects can 
be more easily visualised by displaying the temporally-averaged median values 
across the entire field-of-view. The lower-left panel of Figure~{\ref{images}} 
displays these values normalised to their individual standard 
deviations, $\sigma$. Here, more than 
$63.1$\% ($165{\,}592$~pixels) of the field-of-view display 
negative medians, with only $33.8$\% ($90{\,}870$~pixels) showing 
median values greater than $0$.

The lower-left panel of Figure~{\ref{images}} can be used to 
reveal important information regarding what types of lightcurves 
contribute to those pixels displaying highly negative medians. 
From the definition of this image, it is clear that darker pixels will have 
median values significantly below the mean value of $0$. Therefore, we 
can select the most negative pixels, and display the resulting lightcurves 
to examine why their time-averaged median values are so low. 
The upper panel of Figure~{\ref{lightcurve}} displays a $500$~s 
section of a lightcurve that corresponds to a pixel with a median average of 
$-1.575 \pm 0.002$ when normalised to the standard deviation, or 
$-7.48 \pm 0.07$~DN{\,}s$^{-1}$ in raw units. This lightcurve 
clearly shows a small-scale impulsive event, with a peak rise in 
intensity of $\sim$$60$~DN{\,}s$^{-1}$ above the mean, 
corresponding to a rise of $\sim$$7$\% above the background.
This event is small enough to evade 
our initial intensity threshold of $50$\%, which was designed to filter 
out macroscopic events such as 
H$\alpha$ microflares,
yet large enough to 
substantially diverge the pixel median from the average value as a result of the 
impulsive rise and gradual decay. This example is at the 
extreme end of the median scale. However, smaller impulsive events, 
which may be difficult, if not impossible to detect by eye, may result 
in less severe negative medians. A collection of pseudo-random, 
impulsive events (i.e. nanoflares) that are at, or below the visual detection 
limit, may be the cause of the overall distribution asymmetries 
present in our observations. 

Since our filtering thresholds only removed $\approx$$3.4$\% of 
the total number of pixels, a natural question arises as to what 
contribution the filtered fluctuations would have on the measured 
H$\alpha$ distributions displayed in the upper panels of 
Figure~{\ref{distributions}}. To test the robustness of our 
methodologies, we generated identical distributions for the 
entire observational field-of-view (i.e., including all previously 
discarded pixels). The resulting distributions are overplotted 
in the upper panels of Figure~{\ref{distributions}} using solid 
red lines. Using the entire field-of-view, the observed intensity 
fluctuations remain in close agreement with those obtained 
using the filtered image sequence, implying significant robustness 
in our chosen methodologies. Differences between the respective 
distributions are incredibly subtle, and most likely difficult to 
identify by eye. However, there is a fractional increase in the 
negative offset after including the previously discarded pixels. 
This is most likely a consequence of including more macroscopic 
H$\alpha$ brightenings in the field-of-view, thus causing the 
separation between the mean and median values to become 
more pronounced as a result of the longer decay timescales 
associated with these features 
\citep[$\sim$$3-5$~minutes;][]{Jes10a}. 
Furthermore, the intensity fluctuation 
profile itself is marginally broader when compared to the 
filtered field-of-view. Again, this is likely attributed to the 
inclusion of more rapidly evolving and/or brightening 
structures within the field-of-view, thus causing greater amplitude 
fluctuations to be included in the far wings of the distribution. The 
observed median fluctuations of the filtered and full 
fields-of-view have almost identical negative offsets. 
This is most likely a 
consequence of larger amplitude H$\alpha$ brightenings 
having a reduced occurrence rate when compared to 
nanoflare activity, resulting in the peak median offset being 
determined solely by the relatively frequent nanoflares.
However, as per the intensity fluctuation distribution, the median values 
of the full field-of-view demonstrate a slightly more broadened 
profile when compared to the filtered dataset.  
The often significantly longer decay times associated with 
large-scale chromospheric brightenings 
helps to further separate the statistical mean and median values, thus 
contributing to more negative offsets. Contrarily, rapidly evolving 
chromospheric features, from either spicule-type transverse 
oscillations or H$\alpha$ microflare activity with decay times 
similar to that of a nanoflare ($\sim$$51$~s), will help to 
negate the median offset and thus contribute to 
more positive values.

Interestingly, while the observed intensity fluctuations 
(upper-left panel of Figure~{\ref{distributions}}) are predominantly 
negatively offset, the positive 
tail of the distribution appears to remain elevated beyond that 
of the comparative Gaussian centered about zero. This implies a degree of 
positive skewness to the observed distributions. To quantify this, we 
calculated the Fisher and Pearson coefficients related to the 
distribution, finding values of $0.111$ and $0.096$, respectively. This 
is a slight degree of skewness, but it implies that small-scale 
characteristics embedded 
within the data are promoting a positive skew. Since we set our 
intensity-filtering threshold to $50$\% above the line-of-best-fit, some 
larger impulsive brightenings may still be present in the data 
(see, e.g., the upper panel of Figure~{\ref{lightcurve}}). These 
more-significant impulsive events will result in contributions to 
larger $I/{\sigma}_{P}$ values, thus causing the positive tails 
of the distributions to stay elevated over a wider range. 
To test this theory, we also calculated the Fisher and Pearson 
coefficients for the distributions incorporating the entire 
observational field-of-view (i.e., including all previously 
filtered pixels), with values of $0.115$ and $0.099$ found, respectively. 
These marginally inflated values indicate a higher degree of positive 
skewness when more macroscopic H$\alpha$ brightenings 
are included in the distribution, thus strengthening our 
interpretation. 
Lowering the intensity threshold may help to reduce these 
extended tails (i.e., the skewness). However, since we believe 
that nanoflare activity is of a similar 
magnitude to the Poisson noise statistics, there is a fine line 
between removing very small-scale H-alpha microflare events 
(which may contribute to the skewness) and cropping nanoflare 
activity itself. While the number statistics present in 
\citet{Ter11} are significantly lower than what we present here, 
a degree of positive skew can also be viewed in their intensity 
distributions (see, e.g., Figures~4 \& 6 of \citealt{Ter11}). The 
authors do not attempt to interpret this phenomenon, but 
it is interesting to note that skewness appears to be a feature 
synonymous with both chromospheric and coronal observations.

\subsection{Monte Carlo Simulations}
To investigate further, we performed a series of Monte Carlo simulations. 
Following \citet{Ter11}, we assumed as a null hypothesis that all pixel fluctuations 
are solely due to photon noise 
convolved with an intrinsically flat background. 
To simulate this, we first created a time-averaged H$\alpha$ image 
by averaging all $4040$ individual frames together. The resulting emission 
map is shown in the upper-right panel of Figure~{\ref{images}}, and forms 
the basis of our synthetic time series. A new datacube, $4040$ frames in 
duration, is generated with an identical time-averaged emission map occupying 
each time stamp. Then, we introduce detector noise at each pixel using  
Poisson statistics which have the same average fluctuation amplitudes 
that we observed in the real data (lower-right panel of Figure~{\ref{images}}). 
Finally, we apply the same analysis 
routines to our synthetic dataset, with the resulting intensity fluctuations and 
pixel medians displayed in the middle-left and middle-right panels, 
respectively, of Figure~{\ref{distributions}}. Here, the measured median 
fluctuation is $-0.0010 \pm 0.0002$, while the median average is 
$-0.0027 \pm 0.0008$. Under normal circumstances Poisson 
statistics introduce an degree of asymmetry to a photon-based 
distribution as a result of discrete data sampling. 
However, as the sample size increases, a typical Poisson 
distribution becomes more Gaussian-like, and as a result, more symmetric. 
Our synthetic distributions incorporate in excess of $1.02 \times 10^{9}$ individual 
pixels, and as both measured values are very close to $0$, the resulting 
distributions closely follow the Gaussians of unit width overplotted 
(dashed line) in the middle panels of Figure~{\ref{distributions}}. Thus, 
the large negative asymmetries present in the observations cannot be a 
direct consequence of Poisson statistics alone.

\begin{figure}
\epsscale{1.0}
\plotone{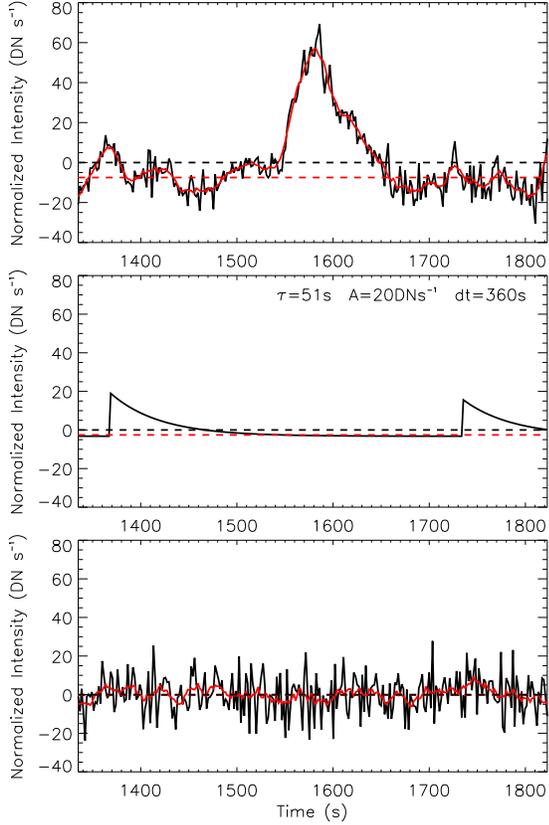}
\caption{A $500$~s duration lightcurve (solid black line; upper), taken from a 
pixel that displays a highly negative median ($-1.575 \pm 0.002$) when 
normalised to the standard deviation. 
Identical $500$~s duration lightcurves, for a Monte Carlo simulation 
with parameters $\tau = 51$~s, $A = 20$~DN{\,}s$^{-1}$, and 
$dt = 360$~s, are displayed without (middle) and with (lower) added 
photon noise. Nine-point ($\approx$$16$~s) 
running averages are displayed using solid red lines, while the dashed 
black and red horizontal lines mark the lightcurve average 
($0$~DN{\,}s$^{-1}$) and median 
values, respectively. A preference for negative medians exists even when the 
impulsive amplitudes are lower than the photon noise. 
The horizontal axes 
displays time from the start of the observing sequence at $17$:$51$~UT.
\label{lightcurve}}
\end{figure}

Next, we introduced a series of impulsive rises in intensity, followed by 
exponential decays, in an attempt to replicate a typical time series 
dominated by nanoflare activity. 
We must stress that the physics of a cooling plasma does not 
necessarily follow a strict exponential decay. In reality, it is a broken power-law 
distribution with different indices for evaporative and non-evaporative processes, in 
addition to whether the plasma is conductively or radiatively cooling \citep{Ant78}. 
However, we chose a more simplistic exponential decay shape to make 
parameterizing the cooling and constraining the decay time more straightforward.
As performed by \citet{Ter11}, we allowed 
the impulsive events to be governed by three distinct parameters: the 
amplitude, $A$, of the impulsive rise; the $e$-folding time, $\tau$, of the 
exponential decay phase; and the average time interval between two 
successive perturbations, $dt$. A number of small-scale impulsive 
events are detectable by eye in pixel lightcurves that display a highly 
negative median (see, e.g., the upper panel of Figure~{\ref{lightcurve}}). 
Measurement 
of eight individual decay phases provides $\tau \approx 51 \pm 14$~s. 
This is consistent with the chromospheric work of \citet{DeJ85}, although 
is much lower than previously used coronal values \citep[$360$~s;][]{Ter11}. 
The higher electron densities found in the chromosphere would lead to 
reduced radiative cooling timescales \citep{Palla90}, hence explaining 
the difference in the value of $\tau$ between chromospheric and coronal 
observations. Alternatively, the different values of $\tau$ could be a direct 
consequence of the spatial resolution. A smaller structure may be expected 
to evolve on faster timescales when compared to a more sizeable feature. 
Therefore, one may expect to 
resolve smaller {\it{and}} faster evolving structures in the high resolution 
chromospheric images compared to those found in XRT observations. 
For the purposes of our simulations, we fix the $e$-folding time 
to equal the value measured in our observations ($\tau = 51$~s). 

One of the larger impulsive events that passed through our data cleaning 
procedures has an amplitude, $A \approx 60$~DN{\,}s$^{-1}$ 
(upper panel of Figure~{\ref{lightcurve}}). This event is comparatively large, 
and results 
in a very low lightcurve median value. As a result, we must choose an 
impulsive amplitude which is significantly below $60$~DN{\,}s$^{-1}$ 
to ensure the average median value over the entire field-of-view is closer 
to the observational measurement of $-0.4172 \pm 0.0008$ 
when normalised to the standard deviation, $\sigma$. As 
nanoflares are believed to be at (or below) the current observational 
detection limit, we can choose an impulsive amplitude similar to 
the standard deviation of our observational time series, which includes 
fluctuations due to both detector readout noise and small-scale solar 
variability. The lower-right image in Figure~{\ref{images}} displays 
the standard deviations for the entire field-of-view, and clearly shows how 
regions surrounding the central sunspot have slightly higher 
standard deviations when compared to those in darker, more distant 
locations of the chromospheric canopy. 
The average standard deviation 
for the entire field-of-view 
($\sigma_{N}$) is $28$~DN{\,}s$^{-1}$,
which is 
considerably higher than the fluctuations solely due to detector readout 
noise 
($\sigma_{d} \approx 4$~DN{\,}s$^{-1}$).
Employing the full range of 
$\sigma_{N}$ values 
($\approx23-39$~DN{\,}s$^{-1}$),
the resulting photon noise can be computed as 
$\sigma_{p}=\sqrt{\sigma_{N}^{2}-\sigma_{d}^{2}}$. This provides a 
photon noise estimate in the range of 
$\sigma_{p}\approx23-39$~DN{\,}s$^{-1}$,
indicating the dominant 
noise contribution arises directly from photon statistics. However, 
in addition to traditional shot noise characteristics, 
we suggest the larger standard deviations found in these locations 
may also be a direct consequence 
of larger magnetic field concentrations in these areas giving rise to 
bigger impulsive events, and hence more intensity variability. 
Thus, we 
select a series of average amplitudes, $A = 10$, $15$, $20$, and 
$25$~DN{\,}s$^{-1}$, and create a 
random-uniform distribution for each amplitude ranging from 
$50$\% to $150$\% ($12.5$ -- $37.5$~DN{\,}s$^{-1}$ in the 
case of $A = 25$~DN{\,}s$^{-1}$), 
which then forms the selection basis of our impulsive intensity rises. 
The final parameter is the average time interval between two 
successive perturbations, $dt$. In order to compare our Monte Carlo 
simulations with those computed by \citet{Ter11}, we selected a 
range of values, $180 \le dt \le 540$~s, where each is separated 
by $60$~s (i.e., $dt = 180$, $240$, $360$, $420$, $480$, and $540$~s). 
For each value, a Poisson distribution centred on the chosen $dt$ is 
generated, which provides a series of successive time intervals between 
adjacent impulsive events. A Poisson distribution is chosen since each 
event will be triggered an integer number of frames after the previous one. 
Impulsive events are then added to the $4040$-frame emission map 
time series, resulting in the average pixel count rates increasing slightly 
as a result of the perturbations having positive values. To ensure 
that individual pixels in the simulated time series have an identical mean 
to that of the actual H$\alpha$ observations, a constant intensity offset is 
applied to each pixel to maintain the same time-averaged 
DN{\,}s$^{-1}$ count rates present in the real data.

The resulting $24$ time series ($\tau = 51$~s; 
$A = 10$, $15$, $20$, $25$~DN{\,}s$^{-1}$; 
$dt = 180$, $240$, $360$, $420$, $480$, $540$~s) 
have non-periodic behaviour, but with low-level 
impulsive events followed by exponential decreasing trends 
(middle panel of Figure~{\ref{lightcurve}}). These 
lightcurves are then subjected to the addition of photon noise according to 
our null-hypothesis test above, and re-analysed using our observational 
routines. An example synthetic lightcurve, with input parameters 
$\tau = 51$~s, $A = 15$~DN{\,}s$^{-1}$, and $dt = 180$~s, is 
displayed in the lower panel of Figure~{\ref{lightcurve}}. 
The resulting intensity fluctuations and 
pixel medians for the $24$ time series were compared to the 
observational measurements, with the closest match occurring for 
the variables $\tau = 51$~s, $A = 20$~DN{\,}s$^{-1}$, and $dt = 360$~s.
The lower-left and lower-right panels of Figure~{\ref{distributions}} display 
the best-match intensity fluctuation and pixel median distributions. 
Here, the measured median 
fluctuation is $-0.1100 \pm 0.0002$, while the median average is 
$-0.4124 \pm 0.0008$.

\section{Concluding Remarks}
\label{conc}

\begin{table*}
\begin{center}
\caption{Comparison between this work and that of \citet{Ter11}. \label{table1}}
\begin{tabular}{lcc}
~&~&~ \\
Parameter			& Current		& \citet{Ter11} 	 	\\
					& Study		&  				\\
\tableline
Observations 								& Optical (H$\alpha$) 	 		& X-Ray (Hinode XRT)	\\
Pixel area									& $512 \times 512$~pixels$^{2}$ 	& $256 \times 256$~pixels$^{2}$ 	\\
Field-of-view 								& $71 \times 71$~arcsec$^{2}$ 	& $256 \times 256$~arcsec$^{2}$ 	\\
Time series duration						& $120$~min 					& $26$~min 	\\
Frames									& $4040$						& $303$		\\
Total pixels								& $1.06 \times 10^{9}$			& $1.99 \times 10^{7}$	\\
Total pixels after data cleaning				& $1.02 \times 10^{9}$			& $1.11 \times 10^{7}$	\\
Median fluctuation / $\sigma_{P}$ (entire FOV)	& $-0.1160 \pm 0.0002$			& not stated	\\
Median / $\sigma$ (entire FOV)				& $-0.4172 \pm 0.0008$			& $-0.0258 \pm 0.0004$	\\
Amplitude ($A$; Monte-Carlo best fit)			& $20$~DN{\,}s$^{-1}$			& $60$~DN{\,}s$^{-1}$	\\
e-folding time ($\tau$; Monte-Carlo best fit)		& $51$~s						& $360$~s	\\
Nanoflare interval ($dt$; Monte-Carlo best fit)		& $360$~s					& $360$~s	\\
~&~&~ \\
\end{tabular}
\footnotesize \\
\end{center}
\end{table*}

The methodology presented in this paper is based on the coronal work of \citet{Ter11}. 
However, our findings have a number of 
key differences, with some quantified in Table~\ref{table1}. First, the nanoflare 
amplitude used in our Monte Carlo simulations produces a substantially 
smaller scatter in the percentage intensity increases that result from the 
simulations of nanoflare activity. For our best-fit case, we use a value 
of $20$~DN{\,}s$^{-1}$, which gives an 
impulsive rise between $1$--$3$\% above the brightest and darkest 
quiescent background pixels, respectively. 
\citet{Ter11} used $A = 60$~DN{\,}s$^{-1}$, which produced a substantially 
wider range of intensity increases, of the order of 
$3$--$200$\% above their brightest ($\sim$$1700$~DN{\,}s$^{-1}$) and 
darkest ($\sim$$30$~DN{\,}s$^{-1}$) quiescent background 
pixels, respectively. This may be a direct consequence of the relatively 
poor signal-to-noise encountered with faint coronal X-ray observations. 
Contrarily, the high signal-to-noise of our HARDcam H$\alpha$ observations 
enables the amplitudes of potential nanoflare activity to be constrained to 
a much narrower window. Importantly, because the nanoflare amplitude is 
only marginally below the observational noise threshold, it seems likely that 
next-generation cameras may be sensitive enough to temporally resolve 
nanoflare activity, especially if contributions due to dark current and 
readout noise can be minimised. Secondly, the average cadence between 
successive nanoflare events that provided the closest resemblance 
to our H$\alpha$ observations is identical to that used by 
\citet{Ter11}. However, while the average cadence, $dt = 360$~s, 
may be identical, the difference in spatial resolution between the two 
studies is drastically different. In our current chromospheric 
simulations, we require a nanoflare event approximately every 
$360$~s over a spatial scale of $\sim$$10{\,}000$~km$^{2}$ ($1$~pixel). 
On the other hand, the coronal simulations of \citet{Ter11} required 
a nanoflare event approximately every $360$~s over a spatial 
scale of $\sim$$525{\,}000$~km$^{2}$ ($1$~pixel). This suggests 
that on comparable spatial scales, there are $\sim$$50$ times more 
nanoflare events in the solar chromosphere compared to the corona. 
Thirdly, our observed distributions of intensity fluctuations and pixel medians 
(upper panels of Figure~{\ref{distributions}}) have a larger negative 
offset than those presented by \citet{Ter11}. Typically, the lower 
impulsive amplitudes used in our Monte Carlo simulations would 
reduce the associated negative offset. However, 
this effect is negated by the much shorter $e$-folding time which 
causes the intensities to drop back down to their quiescent 
value much more abruptly. As a result, the sensitivity to 
small-scale impulsive events in the (optical) chromosphere is substantially 
higher.

Our results suggest that nanoflare activity is readily occurring in the 
solar chromosphere. The energetics associated with these events 
are only fractionally below the noise threshold of our time series, 
and as a result, next-generation instruments with reduced 
readout noise may actually be able to 
temporally resolve such impulsive events.
Even with a relatively small field-of-view size 
($71{\arcsec}\times71{\arcsec}$), our Monte Carlo simulations 
suggest that over $2.5 \times 10^{6}$ impulsive events occur every hour. 
Thus, we suggest that nanoflare heating may be a 
significant heating mechanism in the solar chromosphere.

\acknowledgments
D.B.J. would like to thank STFC for an Ernest Rutherford Fellowship, in addition 
to a dedicated standard grant which allowed this project to be undertaken. 
M.M. is grateful to STFC for research support.

{\it Facilities:} \facility{Dunn (HARDcam)}.

\end{document}